\documentclass[12pt]{iopart}

\expandafter\let\csname equation*\endcsname\undefined
\expandafter\let\csname endequation*\endcsname\undefined

\usepackage{amsmath}
\usepackage{amssymb}
\usepackage{siunitx}
\usepackage{graphicx}
\usepackage{geometry}
\usepackage{xcolor}
\usepackage{placeins}
\usepackage[normalem]{ulem}
\usepackage[%
    style=phys,%
    articletitle=true,biblabel=brackets,%
    chaptertitle=false,pageranges=false%
]{biblatex}
\DeclareFieldFormat
  [article,inbook,incollection,inproceedings,patent,thesis,unpublished]
  {title}{\mkbibemph{#1\isdot}}
\begin{document}

\title[Asymptotic Matching the SCE to Approximate $K_\alpha(x)$]{Asymptotic Matching the Self-Consistent Expansion to Approximate the
Modified Bessel Functions of the Second Kind}

\author{Chanania Steinbock and Eytan Katzav}
\address{Racah Institute of Physics, The Hebrew University, Jerusalem 9190401, Israel}
\ead{chanania.steinbock@mail.huji.ac.il}

\begin{abstract}
The self-consistent expansion (SCE) is a powerful technique for obtaining
perturbative solutions to problems in statistical physics but it suffers
from a subtle problem -- too much freedom! The SCE can be used to
generate an enormous number of approximations but distinguishing the
superb approximations from the deficient ones can only be achieved
after the fact by comparison to experimental or numerical results.
Here, we propose a method of using the SCE to a priori obtain uniform
approximations, namely asymptotic matching. If the asymptotic behaviour
of a problem can be identified, then the approximations generated
by the SCE can be tuned to asymptotically match the desired behaviour
and this can be used to obtain uniform approximations over the entire
domain of consideration, without needing to resort to empirical comparisons.
We demonstrate this method by applying it to the task of obtaining
uniform approximations of the modified Bessel functions of the second
kind, $K_{\alpha}\left(x\right)$.
\end{abstract}


\submitto{\jpa}
\maketitle

\section{Introduction}

Perturbative techniques play a major role in many areas of physics
though such methods are often limited to problems where a well-defined
small parameter can be found \cite{BenderOrszag1999Book,Hinch1991Book}.
In typical many-body systems, such small parameters are often lacking
requiring the application of sophisticated resummation techniques
such as resummation by Pad\'{e} approximant \cite{BenderOrszag1999Book,PadeApproximantsBook1996}
or the infamous renormalisation group \cite{McComb2003}. Though
such techniques can be extremely powerful, there are well known situations
where such resummation techniques simply have no access to the strong
nonlinear coupling regime. For instance, it is well established that
the renormalisation group has no access to the strong coupling regime
of the Kardar-Parisi-Zhang (KPZ) equation describing surface growth
at \emph{any} order \cite{Wiese1998}. 

While perturbative techniques are ubiquitous in physics, applications
to pure mathematics also abound. For instance, characterisation of
the asymptotic behaviour of the so called ``special functions'' has
been a cornerstone of mathematics over the last two centuries \cite{AbramowitzStegunBook,Lebedev1965,Luke1969,NIST:DLMF}.
One of the goals of this paper will be to demonstrate how the techniques
developed by physicists for studying many-body systems can be used
to gain insight into approximating such functions.

The self-consistent expansion (SCE) is a powerful technique in statistical
physics for obtaining perturbative expansions of many-body interacting
systems. First developed by Schwartz and Edwards to investigate the
KPZ equation for surface growth \cite{Schwartz1992}, it has since
been applied to a number of problems in statistical physics, including
generalisations of the KPZ equation \cite{Schwartz1998,Katzav1999,Katzav2002,Schwartz2002,Katzav2002a,Katzav2003a,Katzav2003,Katzav2004a,Katzav2004},
turbulence \cite{Edwards2002} , wetting fronts and fracture \cite{Katzav2006,Katzav2007,Katzav2007b,Katzav2013},
the XY-model \cite{Li1996} and fluctuating elastic sheets \cite{Steinbock2022,Steinbock2023,Steinbock2023b}.
The basic idea of the method is that when performing a perturbative
expansion of any particular system, one will always have various degrees
of freedom in the selection of the zeroth order system. By a posteriori
selecting the zeroth order system in a manner which is self-consistent
with the approximation order, one can obtain superb and even convergent
expansions for systems whose ordinary perturbative expansions are
known to diverge \cite{Schwartz2008,Remez2018}. 

Ironically, one subtle limitation of the SCE is the enormous freedom
in how one self-consistently determines the zeroth order system. In
practice, it seems that convergence of the perturbative expansion
can be obtained by appropriately modifying the self-consistent criteria
simultaneously with the order of the expansion being considered \cite{Schwartz2008,Remez2018}
but there is no a priori way of knowing to what extent the self-consistent
criteria should be modified. As such, the SCE provides the practitioner
with a huge family of approximations but no way of determining which
approximations are excellent and which are deficient. Aside from comparison
with experimental or numerical results, there is little which can
be done to determine the validity of any given approximation.

One approach to resolving this issue is described in \cite{Cohen2016},
where it is suggested that the quality of various sets of approximations
can be assessed by considering their variance, that is, the extent
to which the different approximations vary from each other. A set
consisting of mostly good approximations will have a relatively small
variance while a set containing bad approximations will tend to have
a larger variance as it is unlikely for bad approximations to all
be bad in the same way. As Tolstoy begins Anna Karenina, ``All happy
families are alike; each unhappy family is unhappy in its own way''
\cite{Tolstoy_Anna_Karenina}. This approach adopts another level
of self-consistent reasoning, which is needed when facing very little
or no analytical information about the system. In the very few cases
where a systematic comparison against an external reference can be
made \cite{Cohen2016,Remez2018}, this idea can be verified, however
experience teaches that hard and analytically intractable problems
may challenge such expectations. For instance, conditionally convergent
series show that series can converge but not to the expected value, 
a well-known example being the Taylor expansion of the function
$\exp(-1/x^2)$ around $x = 0$ which is well defined yet converges to $0$ everywhere.

Here, we propose a different approach to this problem, namely asymptotic
matching. It is often the case that the full system under consideration
can be solved in various asymptotic limits \cite{BenderOrszag1999Book}.
If the approximations provided by the SCE can be forced to match up
with these asymptotic solutions, one immediately greatly improves
the likelihood that uniform approximations will be achieved over the
entire domain under investigation.

In this paper, we demonstrate how to apply this method of asymptotic
matching the SCE to obtain novel uniform approximations of the modified
Bessel functions of the second kind. The Bessel functions and their
modified variants are important functions in mathematics and physics,
appearing frequently in the context of wave propagation and oscillations
\cite{NIST:DLMF,Lebedev1965}. Despite this, the series representations
of the modified Bessel function of the second kind $K_{\alpha}\left(x\right)$
are only asymptotic, thereby sharply limiting their range of applicability.
While classical techniques such as hyperasymptotics may be able to
extract insight from the divergent character of these series \cite{SegurAsymptoticsBook,Berry1990,Berry1991,Berry1988,Berry1989}
and novel techniques based on AI are still preliminary \cite{Rabemananjara2021},
series expansions which are manifestly convergent and exhibit uniform
convergence are of substantial value. 
Some recent work has had limited success at deriving an approximation by
imposing the desired asymptotic behaviours to an exponential 
ansatz in an ad hoc manner \cite{Palade2023}.
Here, we show how
the method of asymptotically matching the SCE can be used to systematically obtain
extremely precise uniform approximations of these functions.

While classical asymptotic matching, such as boundary layer theory, involves finding an interval over which both large and small asymptotics hold such that they can be stitched together \cite{BenderOrszag1999Book}, the approach we will use here does not require the determination or identification of such an interval. Instead, conceptually, the SCE will be used to generate families of approximations over the entire positive real axis. From these families, the approximations with the correct large and small asymptotics will be selected. This has the added advantage of making the method extremely intuitive and straight-forward to use.

In Sec.~\ref{sec:Bessel Functions}, we recap the known asymptotic
properties of the modified Bessel functions and describe their limitations.
In Sec.~\ref{sec:SCE}, we demonstrate how the problem of obtaining
uniform approximations of $K_{\alpha}\left(x\right)$ can be converted
into a classic problem in statistical physics of determining the partition
function of a particle trapped in a potential well under thermal equilibrium.
This is followed by an application of the SCE with asymptotic matching
to obtain uniform approximations of $K_{\alpha}\left(x\right)$. In
Sec.~\ref{sec:Results}, the obtained approximations are compared
with the exact result and a discussion of the method and its limitations
appears in Sec.~\ref{sec:Discussion}.

\section{The Modified Bessel Functions\label{sec:Bessel Functions}}

The modified Bessel functions are defined by considering solutions
$y\left(x\right)$ to the second order linear ODE \cite{NIST:DLMF}
\begin{equation}
x^{2}\frac{d^{2}y}{dx^{2}}+x\frac{dy}{dx}-\left(x^{2}+\alpha^{2}\right)y=0\,.
\end{equation}
As a linear second order ODE, the solutions to such an equation are
linearly spanned by two independent solutions, commonly denoted $I_{\alpha}\left(x\right)$
and $K_{\alpha}\left(x\right)$, known as the modified Bessel functions
of the first and second kind respectively. A series solution defines
$I_{\alpha}\left(x\right)$ as
\begin{equation}
I_{\alpha}\left(x\right)=\left(\frac{x}{2}\right)^{\alpha}\sum_{k=0}^{\infty}\frac{1}{4^{k}k!\Gamma\left(\alpha+k+1\right)}x^{2k}\,,\label{eq:I series}
\end{equation}
where $\Gamma\left(z\right)$ denotes the Gamma function \cite{NIST:DLMF}.
In turn, $K_{\alpha}\left(x\right)$ can be defined by the series
solution
\begin{equation}
K_{\alpha}\left(x\right)=\frac{\pi}{2}\frac{I_{-\alpha}\left(x\right)-I_{\alpha}\left(x\right)}{\sin\left(\alpha\pi\right)}\,,\label{eq:K series 1}
\end{equation}
if $\alpha\not\in\mathbb{Z}$, or by
\begin{equation}
K_{n}\left(x\right) = \frac{\left(-1\right)^{n-1}}{2}\left[\left. \frac{\partial I_{\alpha}\left(x\right)}{\partial\alpha} \right|_{\alpha=n}+\left.\frac{\partial I_{\alpha}\left(x\right)}{\partial\alpha}\right|_{\alpha=-n}\right]\,,\label{eq:K series 2}
\end{equation}
if $\alpha=n\in\mathbb{Z}$. Note that $K_\alpha(x)$ diverges as $x$ tends to $0$. In particular, $K_\alpha(x)\sim x^{-\alpha}$ as $x\rightarrow0 $.

\begin{figure}
\begin{centering}
\vspace*{0.4cm}
\includegraphics[width=0.8\textwidth]{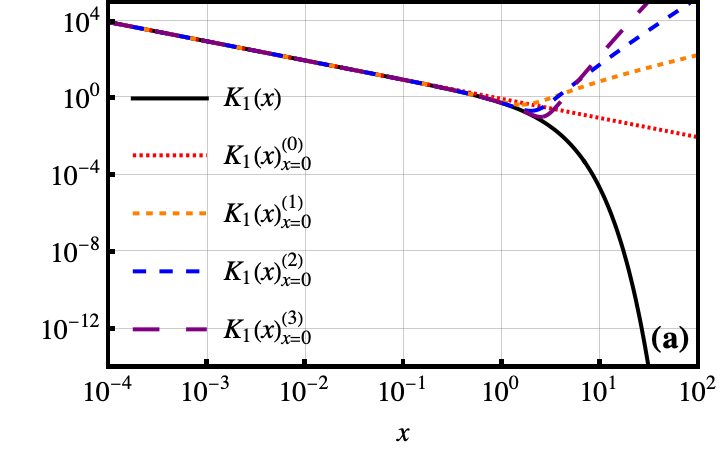}
\par\end{centering}
\vspace{0.2cm}

\begin{centering}
\includegraphics[width=0.8\textwidth]{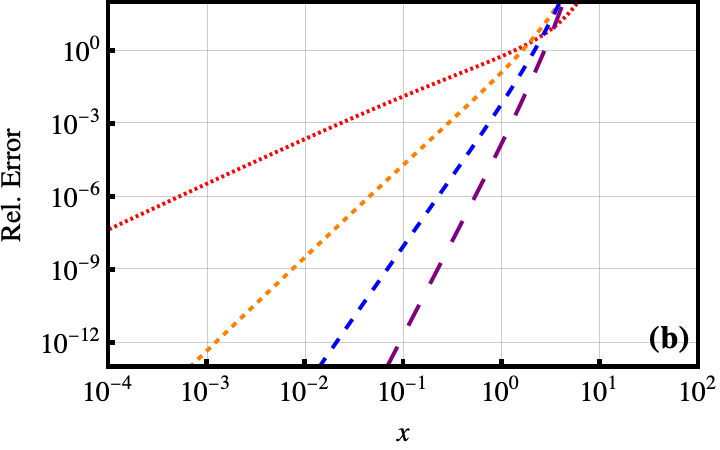}
\par\end{centering}
\caption{\textbf{(a)} The modified Bessel function $K_{1}\left(x\right)$ (solid)
compared with its first few small $x$ approximations, $K_{1}\left(x\right)_{x=0}^{\left(m\right)}$
where $m$ denotes the order at which the series is truncated, given
by Eq.~(\ref{eq:K series 2}) (dashed). \textbf{(b)} The relative
error of each approximation $\left|K_{1}\left(x\right)-K_{1}\left(x\right)_{x=0}^{\left(m\right)}\right|/K_{1}\left(x\right)$.
The approximations become increasingly accurate for small $x$ but
are simply never able to approximate well the large $x$ decaying
tail.\label{fig:Bessel 1 small x}}
\end{figure}

While such series for $K_{\alpha}\left(x\right)$ are useful if $x$
is small, for large values of $x$, $K_{\alpha}\left(x\right)$ is
characterised by a decaying asymptotic behaviour 
\begin{equation}
K_{\alpha}\left(x\right)\xrightarrow{x\rightarrow\infty}\sqrt{\frac{\pi}{2x}}e^{-x}\,.\label{eq:K large x}
\end{equation}
Denoting a truncation of Eqs.~(\ref{eq:K series 1}) or (\ref{eq:K series 2})
after $m$ terms by $K_{\alpha}\left(x\right)_{x=0}^{\left(m\right)}$,
any truncation after a finite number of terms is unable to capture
this decay and thus these series expansions are of limited use for
large $x$. Fig.~\ref{fig:Bessel 1 small x} demonstrates this failure
for a typical value of $\alpha$, namely $\alpha=1$. 

\begin{figure}[t]
\begin{centering}
\vspace*{0.4cm}
\includegraphics[width=0.8\textwidth]{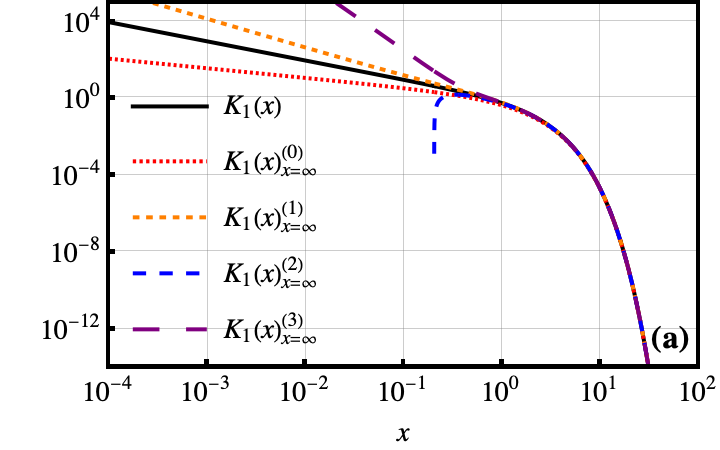}
\par\end{centering}
\vspace{0.2cm}

\begin{centering}
\includegraphics[width=0.8\textwidth]{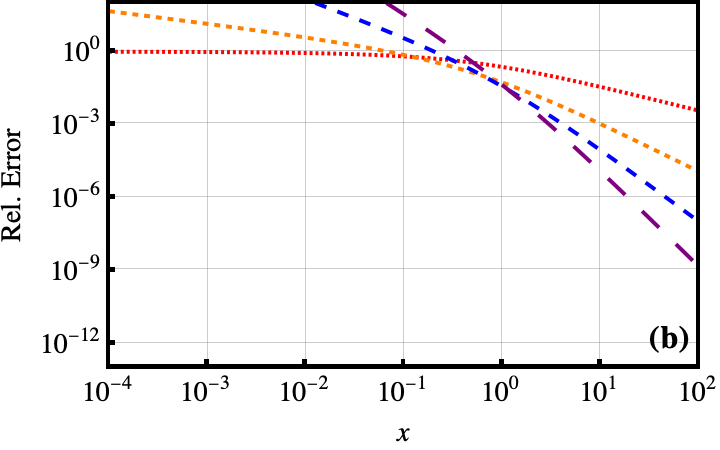}
\par\end{centering}
\caption{\textbf{(a)} The modified Bessel function $K_{1}\left(x\right)$ (solid)
compared with its first few large $x$ approximations, $K_{1}\left(x\right)_{x=\infty}^{\left(m\right)}$
where $m$ denotes the order at which the series is truncated, given
by Eq.~(\ref{eq:K large x series}) (dashed). \textbf{(b)} The relative
error of each approximation $\left|K_{1}\left(x\right)-K_{1}\left(x\right)_{x=\infty}^{\left(m\right)}\right|/K_{1}\left(x\right)$.
The approximations become increasingly accurate for large $x$ but
are simply never able to approximate well the small $x$ behaviour.\label{fig:Bessel 1 big x}}
\end{figure}

More generally, for large values of $x$, $K_{\alpha}\left(x\right)$
has the known asymptotic expansion \cite{NIST:DLMF}
\begin{equation}
K_{\alpha}\left(x\right)\sim\sqrt{\frac{\pi}{2x}}e^{-x}\sum_{k=0}^{\infty}\left(-1\right)^{k}\frac{a_{k}\left(\alpha\right)}{x^{k}}\,,\label{eq:K large x series}
\end{equation}
where
\begin{equation}
a_{k}\left(\alpha\right)=\frac{\left(4\alpha^{2}-1^{2}\right)\left(4\alpha^{2}-3^{2}\right)...\left[4\alpha^{2}-\left(2k-1\right)^{2}\right]}{8^{k}k!}\,,
\end{equation}
however while such an expansion is useful for large $x$, Fig.~\ref{fig:Bessel 1 big x}
shows how truncations after $m$ terms, $K_{\alpha}\left(x\right)_{x=\infty}^{\left(m\right)}$,
are unable to capture the small $x$ behaviour described by Eqs.~(\ref{eq:K series 1})
and (\ref{eq:K series 2}) in the case of $\alpha=1$, which is again
a typical value for $\alpha$. Accordingly, a uniform series expansion,
valid for both large and small $x$ would certainly be of interest.

\FloatBarrier

\section{$K_{\alpha}\left(x\right)$ and Statistical Mechanics\label{sec:SCE}}

In the canonical ensemble of statistical mechanics, a system governed
by Hamiltonian $H\left(u\right)$ in thermal equilibrium with a heat
reservoir at temperature $T$ is fully characterised by its partition
function
\begin{equation}
Z=\int du\,e^{-\beta H\left(u\right)}\,,
\end{equation}
where the integral is over all possible values $u$ can take and $\beta=1/k_{B}T$
($k_{B}$ denotes Boltzmann's constant) \cite{Balescu1975Book,PlischkeBergersen1994Book}.
Further, if $\mathbb{F}$ denotes some observable, the expectation of $\mathbb{F}$,
denoted $\left\langle \mathbb{F}\right\rangle $, is given by
\begin{equation}
\left\langle \mathbb{F}\right\rangle =\frac{1}{Z}\int du\,\mathbb{F}\left(u\right)e^{-\beta H\left(u\right)}\,.
\end{equation}

In the current context, the temperature $T$ will play no role and
thus without loss of generality, we set $\beta=1$ from here on out.
When such integrals cannot be carried out exactly, perturbation theory
provides straight-forward techniques for approximating them. For instance,
suppose that the Hamiltonian $H\left(u\right)$ has the form
\begin{equation}
H\left(u\right)=H_{0}\left(u\right)+\Lambda H_{1}\left(u\right)\,,\label{eq:H0 + H1}
\end{equation}
where $\Lambda$ is nominally small. Then the partition function $Z$
can be expanded in powers of $\Lambda$ by
\begin{equation}
Z=\sum_{n=0}^{\infty}\Lambda^{n}\frac{\left(-1\right)^{n}}{n!}\int du\,\left[H_{1}\left(u\right)\right]^{n}e^{-H_{0}\left(u\right)}\,,\label{eq:Z expansion}
\end{equation}
while the expectation value $\left\langle \mathbb{F}\right\rangle $ can be
expanded in powers of $\Lambda$ by
\begin{equation}
\left\langle \mathbb{F}\right\rangle =\sum_{m=0}^{\infty}\left[\sum_{j=1}^{\infty}\Lambda^{j}\frac{\left(-1\right)^{j+1}}{j!}\left\langle H_{1}^{j}\right\rangle _{0}\right]^{m}\left[\sum_{n=0}^{\infty}\Lambda^{n}\frac{\left(-1\right)^{n}}{n!}\left\langle \mathbb{F} H_{1}^{n}\right\rangle _{0}\right]\,.\label{eq:mean expansion}
\end{equation}
Here, $\left\langle \mathbb{F}\right\rangle _{0}$ denotes an expectation
taken with respect to $H_{0}\left(u\right)$, ie.
\begin{equation}
\left\langle \mathbb{F}\right\rangle _{0}=\frac{\int du\,\mathbb{F}\left(u\right)e^{-H_{0}\left(u\right)}}{\int du\,e^{-H_{0}\left(u\right)}}\,.
\end{equation}
Explicitly, the first few terms in this series expansion for $\left\langle \mathbb{F}\right\rangle $
are
\begin{multline}
\left\langle \mathbb{F}\right\rangle =\left\langle \mathbb{F}\right\rangle _{0}-\Lambda\left[\left\langle \mathbb{F} H_{1}\right\rangle _{0}-\left\langle \mathbb{F}\right\rangle _{0}\left\langle H_{1}\right\rangle _{0}\right]+\\
+\Lambda^{2}\left(\frac{1}{2}\left[\left\langle \mathbb{F} H_{1}^{2}\right\rangle _{0}-\left\langle \mathbb{F}\right\rangle _{0}\left\langle H_{1}^{2}\right\rangle _{0}\right]-\left[\left\langle \mathbb{F} H_{1}\right\rangle _{0}-\left\langle \mathbb{F}\right\rangle _{0}\left\langle H_{1}\right\rangle _{0}\right]\left\langle H_{1}\right\rangle _{0}\right)+O\left(\Lambda^{3}\right)\,.\label{eq:avg expansion}
\end{multline}

Let us now consider the following integral representation of the modified
Bessel function of the second kind \cite{NIST:DLMF}
\begin{equation}
K_{\alpha}\left(x\right)=\int_{0}^{\infty}du\,e^{-x\cosh u}\cosh\left(\alpha u\right)=\frac{1}{2}\int_{-\infty}^{\infty}du\,e^{-x\cosh u-\alpha u}\,.\label{eq:K integral}
\end{equation}
Under the identification
\begin{equation}
H\left(u\right)=x\cosh u+\alpha u\,,\label{eq:Bessel Hamiltonian}
\end{equation}
Eq.~(\ref{eq:K integral}) is the partition function of a system
with a single degree of freedom governed by the Hamiltonian $H\left(u\right)$.
More precisely, the modified Bessel function of the second kind $K_{\alpha}\left(x\right)$
is nothing other than the partition function of an overdamped particle
trapped in an asymmetric hyperbolic potential well in thermal equilibrium.
We will use this observation to obtain uniform series approximations
for $K_{\alpha}\left(x\right)$.

To study this system, we use the self-consistent expansion (SCE).
We begin by considering the zeroth order quadratic Hamiltonian
\begin{equation}
H_{0}\left(u\right)=\frac{\left(u-\xi\right)^{2}}{2\sigma^{2}}+h_{0}\,,
\end{equation}
where $\xi$, $\sigma$ and $h_{0}$ are functions of $x$ and $\alpha$
which denote location, scaling and ``ground state'' parameters and
which are yet to be determined and will be fixed self-consistently
in the continuation. Then we can write our Hamiltonian $H\left(u\right)$
given by Eq.~(\ref{eq:Bessel Hamiltonian}) in the form of Eq.~(\ref{eq:H0 + H1})
by setting
\begin{equation}
H_{1}\left(u\right)=x\cosh u+\alpha u-\frac{\left(u-\xi\right)^{2}}{2\sigma^{2}}-h_{0}
\end{equation}
with $\Lambda=1$. Though in principle, $\Lambda$ should be small,
the self-consistent determination of $\xi$, $\sigma$ and $h_{0}$
will ensure that $\Lambda H_{1}\left(u\right)$ contributes only at
higher orders and thus setting $\Lambda=1$ will pose no problems.
We nevertheless keep $\Lambda$ in our equations for book-keeping
purposes.

According to Eq.~(\ref{eq:Z expansion}), we can expand $K_{\alpha}\left(x\right)$
as

\begin{equation}
K_{\alpha}\left(x\right)=\sqrt{\frac{\pi}{2}\sigma^{2}}e^{-h_{0}}\sum_{n=0}^{\infty}\Lambda^{n}\frac{\left(-1\right)^{n}}{n!}\mathcal{I}_{n}(x,\alpha;\xi,\sigma,h_0)\,,\label{eq:series}
\end{equation}
where we have used the change in variables $v=\left(u-\xi\right)/\sigma$
to simplify the integrals to
\begin{multline}
\mathcal{I}_{n}(x,\alpha;\xi,\sigma,h_0)=\frac{1}{\sqrt{2\pi}} \times\\
\times\int_{-\infty}^{\infty}dv\,e^{-v^{2}/2}\left[x\cosh\left(\sigma v+\xi\right)-\frac{1}{2}v^{2}+\alpha\sigma v+\left(\alpha\xi-h_{0}\right)\right]^{n}\,.
\end{multline}
If the sum in Eq.~(\ref{eq:series}) is truncated such that the highest
power of $\Lambda$ is $m$, we call the resulting expression an $m^{th}$
order approximation of $K_{\alpha}\left(x\right)$ and denote this
by $K_{\alpha}\left(x\right)^{\left(m\right)}$. Though explicit expressions
for general $\mathcal{I}_{n}$ can be written with the aid of the
multinomial theorem, such expressions are overly long and cumbersome
for our purposes. In practice, particular $\mathcal{I}_{n}$ can be
calculated fairly quickly as they are needed. The first few $\mathcal{I}_{n}$
are given by
\begin{align}
\mathcal{I}_{0} & =1\,,\\
\mathcal{I}_{1} & =xe^{\sigma^{2}/2}\cosh\left(\xi\right)+\left(\alpha\xi-h_{0}-\frac{1}{2}\right)\,,\label{eq:I1}\\
\mathcal{I}_{2} & =\frac{1}{2}x^{2}e^{2\sigma^{2}}\cosh\left(2\xi\right)+2x\left[\left(\alpha\xi-h_{0}-\frac{1}{2}\right)-\frac{\sigma^{2}}{2}\right]e^{\sigma^{2}/2}\cosh\left(\xi\right)+\nonumber \\
 & \qquad+2x\alpha\sigma^{2}e^{\sigma^{2}/2}\sinh\left(\xi\right)+\left[\left(\alpha\xi-h_{0}-\frac{1}{2}\right)^{2}+\alpha^{2}\sigma^{2}+\frac{1}{2}x^{2}+\frac{1}{2}\right]\,.
\end{align}

We now turn to the task of selecting $h_{0}$, $\xi$ and $\sigma$.
The SCE prescribes that these parameters be chosen in such a manner
so that zeroth order approximations be in some sense close to the
exact result. For instance, if we set $\mathcal{I}_{1} = 0$ by choosing
\begin{equation}
h_{0}=xe^{\sigma^{2}/2}\cosh\left(\xi\right)+\alpha\xi-\frac{1}{2}\,,
\end{equation}
then it is clear from Eqs.~(\ref{eq:series}) and (\ref{eq:I1})
that the zeroth order approximation $K_{\alpha}\left(x\right)^{\left(0\right)}$,
will be exact up to first order, i.e.
\begin{equation}
K_{\alpha}\left(x\right)^{\left(0\right)}=K_{\alpha}\left(x\right)^{\left(1\right)}\,.
\end{equation}
In general, one can ensure that the first $m$ corrections to $K_{\alpha}\left(x\right)$
vanish by selecting $h_{0}$ such that
\begin{equation}
\sum_{n=1}^{m}\frac{\left(-1\right)^{n}}{n!}\mathcal{I}_{n}=0\label{eq:sum of I_n}
\end{equation}
though we will not need to resort to this here. 

In previous works, the location and scaling parameters, $\xi$ and
$\sigma$, are typically self-consistently determined by demanding
that various moments calculated at lowest order $\left\langle u^{k}\right\rangle _{0}$
be exact. Here, we take a similar approach though with slightly greater
flexibility. To determine $\xi$ and $\sigma$ in a self-consistent
manner, let us try to ensure that the zeroth order \emph{moment generating
function} $\left\langle e^{uz}\right\rangle _{0}$ of our system is
exact up to first order. According to Eq.~(\ref{eq:avg expansion}),
this will be the case if

\begin{equation}
\left\langle e^{uz}H_{1}\right\rangle _{0}-\left\langle e^{uz}\right\rangle _{0}\left\langle H_{1}\right\rangle _{0}=0\,.\label{eq:mgf constraint original}
\end{equation}
Since
\begin{align}
\left\langle e^{uz}\right\rangle _{0} & =e^{\frac{\sigma^{2}}{2}z^{2}+\xi z}\,,\\
\left\langle H_{1}\right\rangle _{0} & =\mathcal{I}_{1}=xe^{\sigma^{2}/2}\cosh\left(\xi\right)+\left(\alpha\xi-h_{0}-\frac{1}{2}\right)\,,\\
\left\langle e^{uz}H_{1}\left(u\right)\right\rangle _{0} & =e^{\frac{\sigma^{2}}{2}z^{2}+\xi z}\Biggl[xe^{\sigma^{2}/2}\cosh\left(\sigma^{2}z+\xi\right)-\frac{\sigma^{2}}{2}z^{2}+\nonumber \\
 & \qquad\qquad\qquad\qquad+\alpha\sigma^{2}z+\left(\alpha\xi-h_{0}-\frac{1}{2}\right)\Biggr]\,,
\end{align}
we find that $\left\langle e^{uz}\right\rangle _{0}$ will be exact
up to first order if
\begin{equation}
0=xe^{\sigma^{2}/2}\left[\cosh\left(\sigma^{2}z+\xi\right)-\cosh\left(\xi\right)\right]-\frac{\sigma^{2}}{2}z^{2}+\alpha\sigma^{2}z\,.\label{eq:mgf constraint}
\end{equation}
As the right-hand side of this equation is a function of $z$, no
particular choice of $\xi$ and $\sigma$ can force it to vanish identically
for all $z$. We can however force the right-hand side to vanish at
two particular points $z_{1}$ and $z_{2}$. Let us suppose we have
selected these points. Using the trigonometric identity
\begin{equation}
\cosh\left(A+B\right)-\cosh\left(B\right)=2\sinh\left(\frac{A}{2}\right)\sinh\left(\frac{A}{2}+B\right)\,,
\end{equation}
we immediately find that
\begin{equation}
\xi=-\sinh^{-1}\left[\frac{\left(2\alpha-z_{i}\right)z_{i}\sigma^{2}e^{-\sigma^{2}/2}}{4x\sinh\left(\frac{\sigma^{2}}{2}z_{i}\right)}\right]-\frac{\sigma^{2}}{2}z_{i}\,.
\end{equation}
where $i=1,2$. In essence, this constitutes two equations for the
two unknowns $\xi$ and $\sigma$ in terms of $z_{1}$ and $z_{2}$.
Given $\sigma$, this equation immediately gives us $\xi$ and since
this equation holds for both $z_{1}$ and $z_{2}$, this equation
also gives us an implicit equation for $\sigma$ 
\begin{multline}
\qquad\sinh^{-1}\left[\frac{\left(2\alpha-z_{1}\right)z_{1}\sigma^{2}e^{-\sigma^{2}/2}}{4x\sinh\left(\frac{\sigma^{2}}{2}z_{1}\right)}\right]+\frac{\sigma^{2}}{2}z_{1}=\\
=\sinh^{-1}\left[\frac{\left(2\alpha-z_{2}\right)z_{2}\sigma^{2}e^{-\sigma^{2}/2}}{4x\sinh\left(\frac{\sigma^{2}}{2}z_{2}\right)}\right]+\frac{\sigma^{2}}{2}z_{2}\,.\qquad
\end{multline}
Applying $\sinh$ to both sides allows us to write this equation without
needing to resort to inverse hyperbolic functions as
\begin{multline}
\frac{\left(2\alpha-z_{1}\right)z_{1}\sigma^{2}}{\tanh\left(\frac{\sigma^{2}}{2}z_{1}\right)}+\sqrt{\left(4x\right)^{2}\sinh^{2}\left(\frac{\sigma^{2}}{2}z_{1}\right)e^{\sigma^{2}}+\left[\left(2\alpha-z_{1}\right)z_{1}\sigma^{2}\right]^{2}}=\\
=\frac{\left(2\alpha-z_{2}\right)z_{2}\sigma^{2}}{\tanh\left(\frac{\sigma^{2}}{2}z_{2}\right)}+\sqrt{\left(4x\right)^{2}\sinh^{2}\left(\frac{\sigma^{2}}{2}z_{2}\right)e^{\sigma^{2}/2}+\left[\left(2\alpha-z_{2}\right)z_{2}\sigma^{2}\right]^{2}}\,.
\end{multline}
We can choose $z_{1}$ and $z_{2}$ however we want. Suppose we choose
some value for $z_{2}$ and then take $z_{2}\rightarrow0$. Then our
expressions for $\xi$ and $\sigma$ simplify substantially and we
obtain
\begin{equation}
\xi=-\sinh^{-1}\left[\frac{\alpha}{x}e^{-\sigma^{2}/2}\right]=-\ln\left[\frac{\alpha}{x}e^{-\sigma^{2}/2}+\sqrt{1+\frac{\alpha^{2}}{x^{2}}e^{-\sigma^{2}}}\right]\label{eq:xi expr}
\end{equation}
and
\begin{equation}
\frac{\left(2\alpha-z_{1}\right)z_{1}\sigma^{2}}{\tanh\left(\frac{\sigma^{2}}{2}z_{1}\right)}+\sqrt{\left(4x\right)^{2}\sinh^{2}\left(\frac{\sigma^{2}}{2}z_{1}\right)e^{\sigma^{2}}+\left[\left(2\alpha-z_{1}\right)z_{1}\sigma^{2}\right]^{2}}=4\alpha\,.
\end{equation}
For practical calculations, it is worth rewriting this equation in
a form which doesn't involve the square root. Ultimately, one can
obtain
\begin{multline}
\left[\left(4x\right)^{2}\sinh^{2}\left(\frac{\sigma^{2}}{2}z_{1}\right)e^{\sigma^{2}}-\left(4\alpha\right)^{2}\right]\sinh^{2}\left(\frac{\sigma^{2}}{2}z_{1}\right)=\\
=\left(2\alpha-z_{1}\right)z_{1}\sigma^{2}\left[\left(2\alpha-z_{1}\right)z_{1}\sigma^{2}-4\alpha\sinh\left(\sigma^{2}z_{1}\right)\right]\,.\label{eq:sigma expr}
\end{multline}

While this equation is still not analytically solvable for $\sigma$,
we can fairly easily determine how $\sigma$ behaves for large and
small values of $x$, under the choice that $z_{1}$ itself doesn't
depend on $x$. In particular, for large $x$, we find that $\sigma^{2}$
decays
\begin{equation}
\sigma^{2}=\frac{1}{x}\left[1-\frac{1}{2x}+\frac{9-2\left(z_{1}^{2}-4\alpha z_{1}+6\alpha^{2}\right)}{24x^{2}}+O\left(x^{-3}\right)\right]
\end{equation}
while for small values of $x$, we find that $\sigma^{2}$ either
tends to a constant or diverges logarithmically depending on the the
choice of $z_{1}$. Explicitly, we have
\begin{equation}
\sigma^{2}\sim\begin{cases}
\sigma_{0}^{2}+O\left(x^{2}\right) & z_{1}<2\alpha\\
-\frac{2}{2\alpha+1}\ln\left(\frac{x}{2\alpha}\right) & z_{1}=2\alpha\\
-\frac{1}{z_{1}+1}\ln\left[-\frac{z_{1}+1}{2\alpha z_{1}\left(z_{1}-2\alpha\right)}\frac{x^{2}}{\ln\left(\frac{z_{1}+1}{2\alpha z_{1}\left(z_{1}-2\alpha\right)}x^{2}\right)}\right] & z_{1}>2\alpha
\end{cases}\,,\label{eq:sigma small x}
\end{equation}
where $\sigma_{0}^{2}$ is determined by the non-trivial solution
to the equation
\begin{equation}
\left(2\alpha-z_{1}\right)z_{1}\sigma_{0}^{2}=2\alpha\left(1-e^{-\sigma_{0}^{2}z_{1}}\right)\,,\label{eq:sigma0 z1 relationship}
\end{equation}
which can be written in terms of the Lambert-$W$ function \cite{NIST:DLMF}
as
\begin{equation}
\sigma_{0}^{2}=\frac{1}{z_{1}}\left[W\left(-\frac{2\alpha}{2\alpha-z_{1}}\exp\left(-\frac{2\alpha}{2\alpha-z_{1}}\right)\right)+\frac{2\alpha}{2\alpha-z_{1}}\right]\,.\label{eq:sigma0 original}
\end{equation}
Note that for $0<z_{1}<2\alpha$, the primary branch $W_{0}$ is used
while for $z_1<0$, the secondary branch $W_{-1}$ must be used instead.

With these expressions in hand, the large $x$ behaviour of $\xi$
and $h_{0}$ immediately follow as
\begin{equation}
\xi=-\frac{\alpha}{x}\left[1-\frac{1}{2x}-\frac{4\alpha^{2}-9}{24x^{2}}+\frac{z_{1}^{2}-4\alpha z_{1}+12\alpha^{2}-8}{24x^{3}}+O\left(x^{-4}\right)\right]
\end{equation}
and
\begin{equation}
h_{0}=x\left[1-\frac{4\alpha^{2}+1}{8x^{2}}-\frac{z_{1}^{2}-4\alpha z_{1}-2}{24x^{3}}+O\left(x^{-4}\right)\right]\,,
\end{equation}
while their small $x$ behaviour follows as
\begin{equation}
\xi\sim\ln\left(\frac{x}{2\alpha}\right)+\begin{cases}
\frac{1}{2}\sigma_{0}^{2} & z_{1}<2\alpha\\
-\frac{1}{2\alpha+1}\ln\left(\frac{x}{2\alpha}\right) & z_{1}=2\alpha\\
-\frac{1}{2\left(z_{1}+1\right)}\ln\left[-\frac{z_{1}+1}{2\alpha z_{1}\left(z_{1}-2\alpha\right)}\frac{x^{2}}{\ln\left(\frac{z_{1}+1}{2\alpha z_{1}\left(z_{1}-2\alpha\right)}x^{2}\right)}\right] & z_{1}>2\alpha
\end{cases}\label{eq:xi small x}
\end{equation}
and
\begin{equation}
h_{0}\sim\alpha\ln\left(\frac{x}{2\alpha}\right)+\alpha-\frac{1}{2}-\begin{cases}
-\frac{1}{2}\alpha\sigma_{0}^{2} & z_{1}<2\alpha\\
\frac{\alpha}{2\alpha+1}\ln\left(\frac{x}{2\alpha}\right) & z_{1}=2\alpha\\
\frac{\alpha}{2\left(z_{1}+1\right)}\ln\left[-\frac{z_{1}+1}{2\alpha z_{1}\left(z_{1}-2\alpha\right)}\frac{x^{2}}{\ln\left(\frac{z_{1}+1}{2\alpha z_{1}\left(z_{1}-2\alpha\right)}x^{2}\right)}\right] & z_{1}>2\alpha
\end{cases}\,.\label{eq:h small x}
\end{equation}
These expressions are sufficient to determine the leading order behaviour
of our approximations for $K_{\alpha}\left(x\right)$.

Beginning with the large $x$ behaviour, we find that the integrals
$\mathcal{I}_{n}$ behave as
\begin{equation}
\mathcal{I}_{n}\sim C_{n}\frac{1}{x^{n}}\,,
\end{equation}
where the $C_{n}$ are just numbers which happen to be given by
\begin{align}
C_{n} & =\left(-\frac{1}{4}\right)^{n}\sum_{\ell=0}^{n}\binom{n}{\ell}\left(-\frac{3}{2}\right)^{\ell}\sum_{j=0}^{2\ell}\binom{2\ell}{j}\frac{\left(2j\right)!}{j!}\left(-\frac{1}{6}\right)^{j}\,,\\
 & =i\sqrt{\frac{3}{2}}\left(-\frac{1}{4}\right)^{n}\sum_{\ell=0}^{n}\binom{n}{\ell}\left(-\frac{3}{2}\right)^{\ell}U\left(\frac{1}{2},2\ell+\frac{3}{2},-\frac{3}{2}\right)\,.
\end{align}
In the latter expression, $U\left(a,b,z\right)$ denotes the confluent
hypergeometric function \cite{NIST:DLMF} and note that despite the
presence of the imaginary factor $i$, the $C_{n}$ are completely
real. Accordingly, for large $x$, the leading order behaviour of
$K_{\alpha}\left(x\right)^{\left(m\right)}$ is just given by the
pre-factor of Eq.~(\ref{eq:series}) and thus we find that at \emph{any}
order $m$
\begin{equation}
K_{\alpha}\left(x\right)^{\left(m\right)}\sim\sqrt{\frac{\pi}{2}\sigma^{2}}e^{-h_{0}}\sim\sqrt{\frac{\pi}{2x}}e^{-x}.
\end{equation}
Comparing this result with Eq.~(\ref{eq:K large x}), we find that
at the very least, our approximations capture the leading order large
$x$ asymptotic behaviour of $K_{\alpha}\left(x\right)$.

For small $x$, matters are more complicated as the leading order
behaviour depends on our choice of $z_{1}$ and all the $\mathcal{I}_{n}$
also contribute at leading order such that the leading order behaviour
of $K_{\alpha}\left(x\right)^{\left(m\right)}$ depends on both $m$
and $z_{1}$. For $z_{1}<2\alpha$ however, we can write
\begin{equation}
K_{\alpha}\left(x\right)^{\left(m\right)}\sim\sqrt{\frac{\pi}{2}\sigma_{0}^{2}}\left(\frac{2\alpha}{x}\right)^{\alpha}e^{-\frac{1}{2}\alpha\sigma_{0}^{2}-\alpha+\frac{1}{2}}\left[1+\sum_{n=2}^{m}\frac{\left(-1\right)^{n}}{n!}\mathcal{I}_{n}\right]\,,
\end{equation}
where we have made use of the fact that our choice of $h_{0}$ ensures
that $\mathcal{I}_{1}=0$. On the other hand, according to Eqs.~(\ref{eq:K series 1})
and (\ref{eq:K series 2}), $K_{\alpha}\left(x\right)$ has the leading
order small $x$ behaviour
\begin{equation}
K_{\alpha}\left(x\right)\sim\frac{1}{2}\Gamma\left(\alpha\right)\left(\frac{2}{x}\right)^{\alpha}\,.
\end{equation}
We thus find that our approximations will match the asymptotic small
$x$ behaviour if $z_{1}$ and $m$ are selected such that
\begin{equation}
\frac{1}{2}\Gamma\left(\alpha\right)=\sqrt{\frac{\pi}{2}\sigma_{0}^{2}}\alpha^{\alpha}e^{-\frac{1}{2}\alpha\sigma_{0}^{2}-\alpha+\frac{1}{2}}\left[1+\sum_{n=2}^{m}\frac{\left(-1\right)^{n}}{n!}\mathcal{I}_{n}\left(x\rightarrow0\right)\right]\,.
\end{equation}
For any approximation order $m$, we can use this equation to determine
$z_{1}$. If there exists a $z_{1}$ which satisfies this equation,
then the approximation $K_{\alpha}\left(x\right)^{\left(m\right)}$
with this value of $z_{1}$ will be asymptotically valid for both
large and small $x$ and thus will describe a uniform approximation
of $K_{\alpha}\left(x\right)$ over all values of $x$. It is important
to appreciate however that there is no guarantee that this equation
will have a solution for $z_{1}$. In such cases, higher order terms
are needed.

Before moving on to the results, it is worth tying up a few loose ends.
First, note that the correct small $x$ asymptotic behaviour is only
obtained for $z_{1}<2\alpha$. For $z_{1}\ge2\alpha$, the additional
logarithmic corrections which enter $\sigma^{2}$, $\xi$ and $h_{0}$,
as described in Eqs.~(\ref{eq:sigma small x}), (\ref{eq:xi small x})
and (\ref{eq:h small x}), introduce additional $x$ dependence which
deviates from the correct asymptotic behaviour. Accordingly, only
the first lines of these equations are relevant. Second, above, we
set $z_{2}\rightarrow0$ almost arbitrarily and it turned out that
this was sufficient to guarantee that the large $x$ behaviour of
our approximations are asymptotically correct. Had this failed, a
more general though less analytically tractable approach could have
been taken in which both $z_{1}$ and $z_{2}$ are simultaneously
varied to asymptotically match our approximations with the exact result.
Additionally, there is another subtlety worth pointing out here. Eq.~(\ref{eq:mgf constraint})
is satisfied identically for $z=0$ thus it would seem that setting
$z_{2}=0$ provides no additional constraint on $\left\langle e^{uz}\right\rangle $.
Despite this, Eqs.~(\ref{eq:xi expr}) and (\ref{eq:sigma expr})
are clearly not tautologies, even after taking the limit $z_{2}\rightarrow0$.
To understand this, we need to reexamine Eq.~(\ref{eq:mgf constraint original}).
Trivially, this equation is satisfied for $z=0$ which is simply a
statement that the normalisation of the underlying probability distribution
is already exact up to first order. Indeed, by our definitions, any
zeroth order Hamiltonian will result in a properly normalised underlying
distribution and thus the normalisation will be exact at \emph{all}
orders. On the other hand, if we differentiate Eq.~(\ref{eq:mgf constraint original})
$k$ times with respect to $z$ and substitute in $z=0$, we obtain
\begin{equation}
\left\langle u^{k}H_{1}\right\rangle _{0}-\left\langle u^{k}\right\rangle _{0}\left\langle H_{1}\right\rangle _{0}=0\,,\label{eq:moment constraints}
\end{equation}
i.e. Eq.~(\ref{eq:mgf constraint original}) can be used to obtain
constraints on the moments $\left\langle u^{k}\right\rangle $ of
our system. Comparing this constraint for $k=1$ with our results
described above, we find that taking $z_{2}\rightarrow0$ is equivalent
to imposing
\begin{equation}
\left\langle uH_{1}\right\rangle _{0}-\left\langle u\right\rangle _{0}\left\langle H_{1}\right\rangle _{0}=0
\end{equation}
on our system, that is, taking $z_{2}\rightarrow0$ is equivalent
to demanding that not only Eq.~(\ref{eq:mgf constraint}) be satisfied
at $z=0$ but also its first derivative and this is equivalent to
imposing that the zeroth order expectation of the mean $\left\langle u\right\rangle _{0}$
be exact up to first order. This explains why setting $z_{2}\rightarrow0$
is nontrivial and goes someway to accounting for its success at capturing
the large $x$ asymptotic behaviour, as imposing that the zeroth order
first moment or ``centre of mass'' be exact up to first order is a
reasonable preliminary expectation of any physically motivated theory.

\section{Results\label{sec:Results}}

Let's simplify and recap our main findings. We expect that the modified
Bessel function of the second kind $K_{\alpha}\left(x\right)$ can
be approximated up to $m^{th}$ order by
\begin{multline}
K_{\alpha}\left(x\right)^{\left(m\right)}=\sqrt{\frac{\pi}{2}\sigma^{2}}\left(\frac{\alpha}{x}\right)^{\alpha}\left(1+\sqrt{1+\frac{x^{2}}{\alpha^{2}}e^{\sigma^{2}}}\right)^{\alpha}\times\\
\times e^{\frac{1}{2}-\frac{1}{2}\alpha\sigma^{2}-\sqrt{\alpha^{2}+x^{2}e^{\sigma^{2}}}}\left[1+\sum_{n=2}^{m}\frac{\left(-1\right)^{n}}{n!}\mathcal{I}_{n}\left(x,\alpha\right)\right]\,,\label{eq:Bessel approx}
\end{multline}
where the $\mathcal{I}_{n}$ are integrals given by
\begin{multline}
\mathcal{I}_{n}\left(x,\alpha\right)=\frac{1}{\sqrt{2\pi}}\int_{-\infty}^{\infty}dv\,e^{-v^{2}/2}\Biggl[\sqrt{\alpha^{2}+x^{2}e^{\sigma^{2}}}\left(e^{-\sigma^{2}/2}\cosh\left(\sigma v\right)-1\right)+\\
-\alpha e^{-\sigma^{2}/2}\sinh\left(\sigma v\right)-\frac{1}{2}v^{2}+\alpha\sigma v+\frac{1}{2}\Biggr]^{n}
\end{multline}
and $\sigma$ is a function of $x$ and $\alpha$ defined implicitly
by the equation
\begin{multline}
\left[\left(4x\right)^{2}\sinh^{2}\left(\frac{\sigma^{2}}{2}z_{1}\right)e^{\sigma^{2}}-\left(4\alpha\right)^{2}\right]\sinh^{2}\left(\frac{\sigma^{2}}{2}z_{1}\right)=\\
=\left(2\alpha-z_{1}\right)z_{1}\sigma^{2}\left[\left(2\alpha-z_{1}\right)z_{1}\sigma^{2}-4\alpha\sinh\left(\sigma^{2}z_{1}\right)\right]\,.\label{eq:sigma eq}
\end{multline}
To solve this equation for $\sigma$, a choice needs to be made for
$z_{1}$ and if possible, it should be chosen so as to ensure that
\begin{equation}
\Gamma\left(\alpha\right)=\sqrt{2\pi\sigma_{0}^{2}}\alpha^{\alpha}e^{-\frac{1}{2}\alpha\sigma_{0}^{2}-\alpha+\frac{1}{2}}\left[1+\sum_{n=2}^{m}\frac{\left(-1\right)^{n}}{n!}\mathcal{I}_{n}\left(x=0,\alpha\right)\right]\,.\label{eq:z equ}
\end{equation}
Here, $\sigma_{0}$ simply denotes $\sigma\left(x\rightarrow0\right)$
and can be written explicitly in terms of the Lambert-$W$ function
as
\begin{equation}
\sigma_{0}^{2}=\frac{1}{z_{1}}\left[W_{-1}\left(-\frac{2\alpha}{2\alpha-z_{1}}\exp\left(-\frac{2\alpha}{2\alpha-z_{1}}\right)\right)+\frac{2\alpha}{2\alpha-z_{1}}\right]\,.\label{eq:sigma0}
\end{equation}
(This is just Eq.~(\ref{eq:sigma0 original}) with a specific choice
of the branch of $W$.)

\begin{figure}
\begin{centering}
\includegraphics[width=1\textwidth]{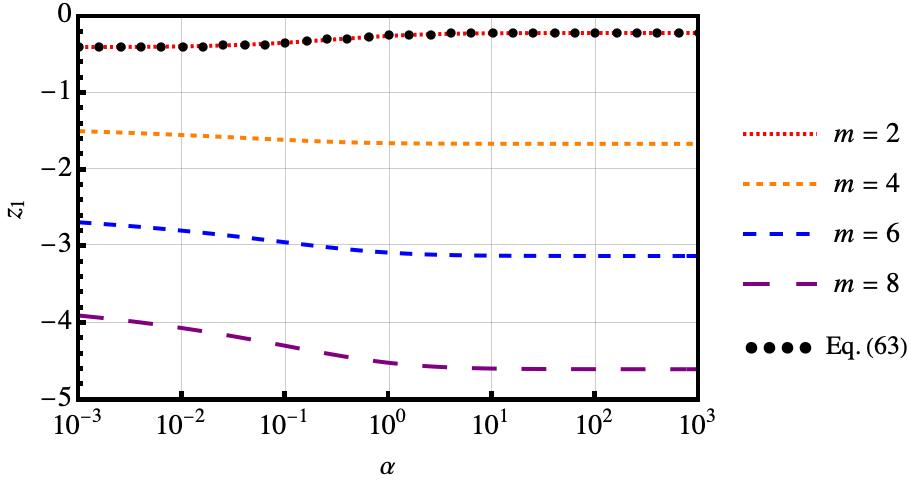}
\par\end{centering}
\caption{$z_{1}$ as a function of $\alpha$, as determined by Eq.~(\ref{eq:z equ}).
For each value of $m$, $z_{1}\left(\alpha\right)$ can be seen to
be a slowly varying sigmoid function over 6 orders magnitude of $\alpha$.
The black dots indicates the Pad\'{e} approximant given by Eq.~(\ref{eq:z1 Pade})
and can be seen to closely approximate the dashed red line corresponding
to the $m=2$ case. \label{fig:z_vs_alpha}}
\end{figure}

For instance, suppose we are interested in the second order approximation
\mbox{($m=2$)}. Since 
\begin{equation}
\mathcal{I}_{2}=\frac{1}{2}\left[x^{2}\left(e^{\sigma^{2}}-1\right)^{2}+2\alpha^{2}\left(e^{\sigma^{2}}-1\right)+1\right]-\sigma^{2}\left[\alpha^{2}+\sqrt{\alpha^{2}+x^{2}e^{\sigma^{2}}}\right]\,,
\end{equation}
Eq.~(\ref{eq:z equ}) requires that we select $z_{1}$ such that
\begin{equation}
\Gamma\left(\alpha\right)=\sqrt{2\pi\sigma_{0}^{2}}\alpha^{\alpha}e^{-\frac{1}{2}\alpha\sigma_{0}^{2}-\alpha+\frac{1}{2}}\left[\frac{5}{4}+\frac{1}{2}\alpha^{2}\left(e^{\sigma_{0}^{2}}-1\right)-\frac{1}{2}\left(\alpha+1\right)\alpha\sigma_{0}^{2}\right]\,,
\end{equation}
where $\sigma_{0}$ is given in terms of $\alpha$ and $z_{1}$ by
Eq.~(\ref{eq:sigma0}). For any given $\alpha$, this equation can
be solved numerically however it is worth appreciating that
its solution for $z_{1}$ as a function of $\alpha$ is just some
sort of sigmoid function. In particular, approximating the gamma function
with Stirling's formula, one can obtain a simple Pad\'{e} approximation
\cite{PadeApproximantsBook1996} of this function 
\begin{equation}
z_{1}\left(\alpha\right)=-\frac{396949+892620\alpha}{1012625+4284576\alpha}\,,\label{eq:z1 Pade}
\end{equation}
which well approximates $z_{1}$ for all $\alpha\ge10^{-3}$, as can
be seen in Fig.~\ref{fig:z_vs_alpha}. For large $\alpha$, $z_{1}\left(\alpha\right)$
just tends to the constant $-5/24$.

Indeed, as shown in Fig.~\ref{fig:z_vs_alpha}, at any order $m$,
whenever Eq.~(\ref{eq:z equ}) has a solution for $z_{1}$, it tends
to be some sort of sigmoid function, varying between two distinct
values for large and small $\alpha$. As these values turn out to
be negative, only the first line of Eqs.~(\ref{eq:sigma small x}),
(\ref{eq:xi small x}) and (\ref{eq:h small x}), corresponding to
$z_{1}<2\alpha$, are ever relevant, which is consistent with the
previously noted requirement for asymptotic matching to be possible.
This also justifies the explicit use of the branch $W_{-1}$ in Eq.~(\ref{eq:sigma0}).
With this function $z_{1}\left(\alpha\right)$ in hand, Eq.~(\ref{eq:sigma eq})
can be used to numerically obtain the function $\sigma\left(x,\alpha\right)$
which can then be substituted into Eq.~(\ref{eq:Bessel approx})
to obtain an $m^{th}$ order approximation for $K_{\alpha}\left(x\right)$.%

Since solving Eq.~(\ref{eq:z equ}) for $z_{1}\left(\alpha\right)$ and Eq.~(\ref{eq:sigma eq}) for 
$\sigma\left(x,\alpha\right)$ entails finding the solutions to highly nonlinear equations, for practical purposes, it is worth briefly outlining robust numerical methods to achieve this. To determine $z_{1}\left(\alpha\right)$, it is in fact more convenient to use Eq.~(\ref{eq:z equ}) to define a nonlinear function of $\sigma_{0}=\sigma\left(x\rightarrow0,\alpha\right)$
\begin{equation}
f\left(\sigma_{0}\right)=\frac{\sqrt{2\pi\sigma_{0}^{2}}\alpha^{\alpha}e^{-\frac{1}{2}\alpha\sigma_{0}^{2}-\alpha+\frac{1}{2}}}{\Gamma\left(\alpha\right)}-\frac{1}{1+\sum_{n=2}^{m}\frac{\left(-1\right)^{n}}{n!}\mathcal{I}_{n}\left(x=0,\alpha\right)}\,,
\end{equation}
such that the solution to Eq.~(\ref{eq:z equ}) is simply the root of this function. The advantage of defining $f\left(\sigma_{0}\right)$ in this manner is that for any $\alpha$ and $m$, $f\left(\sigma_{0}\right)$ is simply a sigmoid function in $\sigma_{0}$ and thus never grows so large that numerical methods break down. Indeed, any ordinary bisection method, starting on the interval $\left[0,1/\alpha\right]$ can be used to rapidly and accurately obtain the root of $f\left(\sigma_{0}\right)$. With $\sigma_{0}\left(\alpha\right)$ in hand, one can now use the relationship given by Eq.~(\ref{eq:sigma0 z1 relationship}) to find a solution for $z_{1}\left(\alpha\right)$. By approximating the function
\begin{equation}
g\left(z_{1}\right)=\sigma_{0}^{2}z_{1}\left(2\alpha-z_{1}\right)-2\alpha\left(1-e^{-\sigma_{0}^{2}z_{1}}\right)\,,\label{eq:g(z1)}
\end{equation}
as a cubic polynomial around $z_{1}=0$,
\begin{equation}
g\left(z_{1}\right)\approx-\left[\frac{1}{3}\alpha\sigma_{0}^{4}z_{1}+\left(1-\alpha\sigma_{0}^{2}\right)\right]\sigma_{0}^{2}z_{1}^{2}\,,
\end{equation}
it immediately becomes apparent that the nontrivial root of $g\left(z_{1}\right)$ must lie inside the interval 
$\left[-3\left(1-\alpha\sigma_{0}^{2}\right)/\left(\alpha\sigma_{0}^{4}\right),\alpha+W_{-1}\left(-\alpha\sigma_{0}^{2}e^{-\alpha\sigma_{0}^{2}}\right)/\sigma_{0}^{2}\right]$ and thus here too, an ordinary bisection method can be used to reliably obtain $z_{1}\left(\alpha\right)$ with any level of precision desired.

Once $z_{1}\left(\alpha\right)$ and $\sigma_0(\alpha)$ have been obtained, either using the method just described or the approximation given in Eq.~(\ref{eq:z1 Pade}), which is valid for $m=2$, together with Eq.~(\ref{eq:sigma0}), the function $\sigma\left(x,\alpha\right)$ needs to be obtained from Eq.~(\ref{eq:sigma eq}). Though we cannot write an explicit expression for $\sigma\left(x,\alpha\right)$ in terms of $x$, we can trivially write the inverse relationship 
\begin{equation}
x=\sqrt{\frac{\alpha^{2}e^{-\sigma^{2}}}{\sinh^{2}\left(\frac{\sigma^{2}}{2}z_{1}\right)}+\frac{e^{-\sigma^{2}}\left(2\alpha-z_{1}\right)z_{1}\sigma^{2}}{16\sinh^{4}\left(\frac{\sigma^{2}}{2}z_{1}\right)}\left[\left(2\alpha-z_{1}\right)z_{1}\sigma^{2}-4\alpha\sinh\left(\sigma^{2}z_{1}\right)\right]}\,.\label{eq:x(sigma)}
\end{equation}
Plotting this function over the domain $\sigma\in(0,\sigma_{0}]$, one finds that this is a monotonically decreasing function of $\sigma$ and is thus invertible. Accordingly, it is easy to determine $\sigma\left(x,\alpha\right)$ by using this equation to calculate $x$ as a function of $\sigma$ and then inverting the result. Note that this inversion does not require solving an equation (not even algebraic).

\begin{figure}
\begin{centering}
\includegraphics[width=0.8\textwidth]{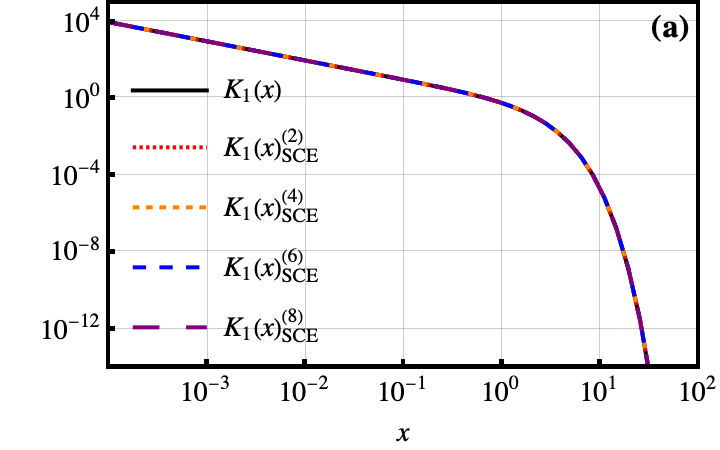}
\par\end{centering}
\vspace{0.5cm}

\begin{centering}
\includegraphics[width=0.8\textwidth]{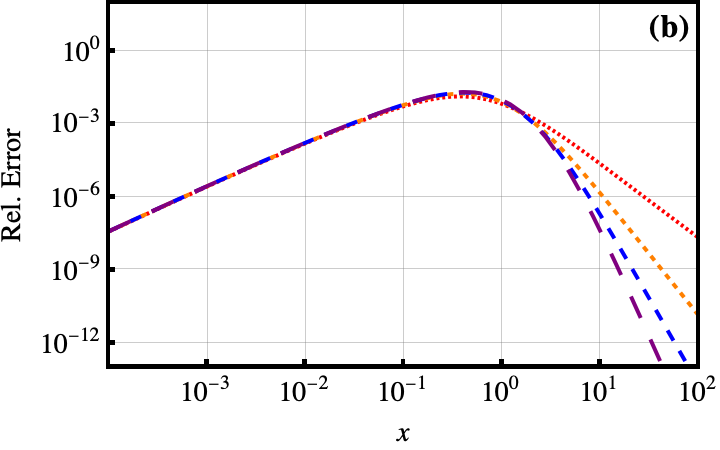}
\par\end{centering}
\caption{\textbf{(a)} The modified Bessel function $K_{1}\left(x\right)$ (solid)
compared with the first few approximations given by the SCE, $K_{1}\left(x\right)_{\text{SCE}}^{\left(m\right)}$
(dashed). \textbf{(b)} The relative error of each approximation. The
approximations are uniformly accurate over all orders of magnitude.\label{fig:Bessel 1 SCE}}
\end{figure}

\begin{figure}
\begin{centering}
\includegraphics[width=1\textwidth]{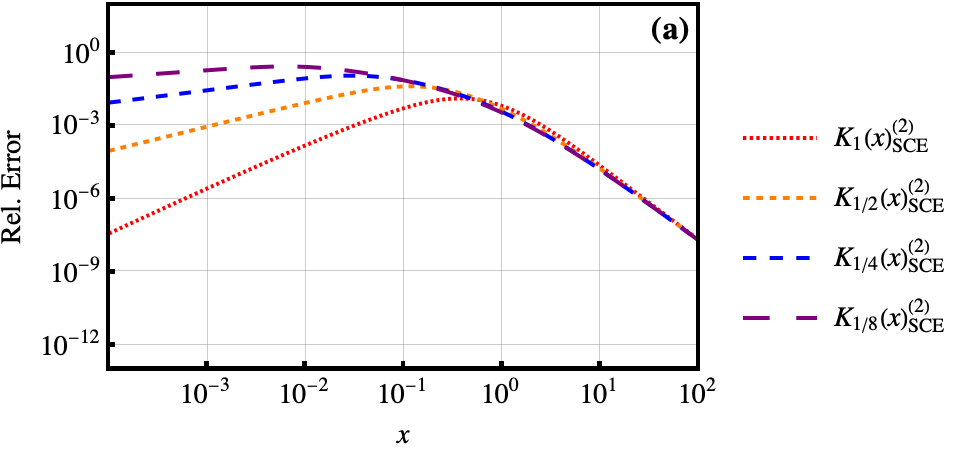}
\par\end{centering}
\vspace{0.5cm}

\begin{centering}
\includegraphics[width=1\textwidth]{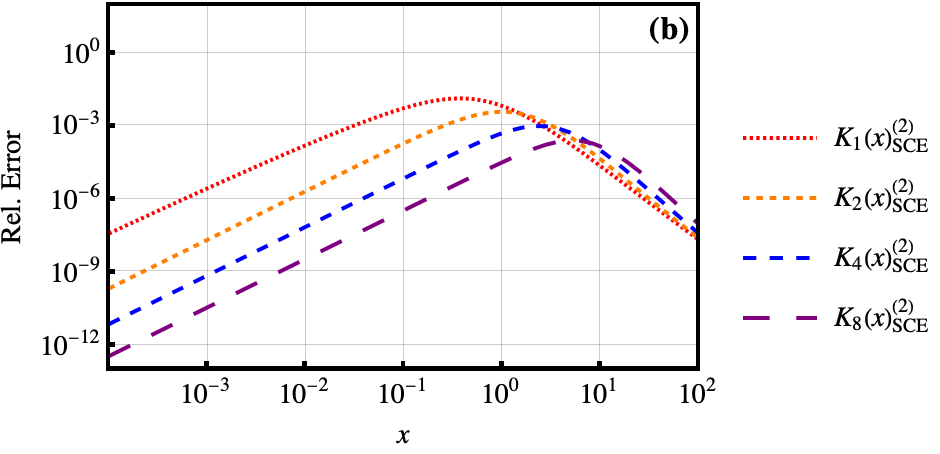}
\par\end{centering}
\caption{The relative error of the second order SCE approximations $K_{\alpha}\left(x\right)_{\text{SCE}}^{\left(2\right)}$.\textbf{
(a)}~$\alpha\in\left\{ 1,1/2,1/4,1/8\right\} $. \textbf{(b)}~$\alpha\in\left\{ 1,2,4,8\right\} $.
The SCE is able to approximate $K_{\alpha}\left(x\right)$ for any
$\alpha > 0$ though it performs substantially better for larger $\alpha$.
\label{fig:Bessel alpha SCE}}
\end{figure}

Fig.~\ref{fig:Bessel 1 SCE} compares the first few of these approximations
for $K_{1}\left(x\right)$ with the exact function. In contrast with
Figs.~\ref{fig:Bessel 1 small x}(a) and \ref{fig:Bessel 1 big x}(a),
Fig.~\ref{fig:Bessel 1 SCE}(a) shows that these approximations accurately
capture the behaviour of $K_{1}\left(x\right)$ over all orders of
magnitude! Indeed, as can be seen from Fig.~\ref{fig:Bessel 1 SCE}(b),
the maximum error of the approximations anywhere is roughly 1\%. Furthermore,
for large $x$, even the lowest order approximation, $K_{1}\left(x\right)_{\text{SCE}}^{\left(2\right)}$,
is more precise than the first few asymptotic expansions shown by
Fig.~\ref{fig:Bessel 1 big x}(b) and higher order approximations
rapidly outclass the asymptotic expansions. On the other hand, for
small $x$, while the correct asymptotic behaviour of our approximations
is achieved by design, the error in the approximation is not necessarily reduced
by taking higher orders. Accordingly, for small $x$, the lowest order
approximation shown $K_{1}\left(x\right)_{\text{SCE}}^{\left(2\right)}$
is as good as the highest and it is reasonable to conjecture that
this behaviour extends to all higher orders. This indicates that unlike
the case for large $x$, asymptotically matching the SCE for small
$x$ is only sufficient to capture the leading order behaviour. Higher
order corrections are not automatically matched as appears to be the
case for large $x$.

It is also worth mentioning that we have only showed even order approximations
of the form $K_{1}\left(x\right)_{\text{SCE}}^{\left(m\right)}$ because
for odd $m$, Eq.~(\ref{eq:z equ}) simply does not have a real solution
for $z_{1}$ and thus small $x$ asymptotic matching is impossible.

\begin{figure}
\begin{centering}
\includegraphics[width=0.8\textwidth]{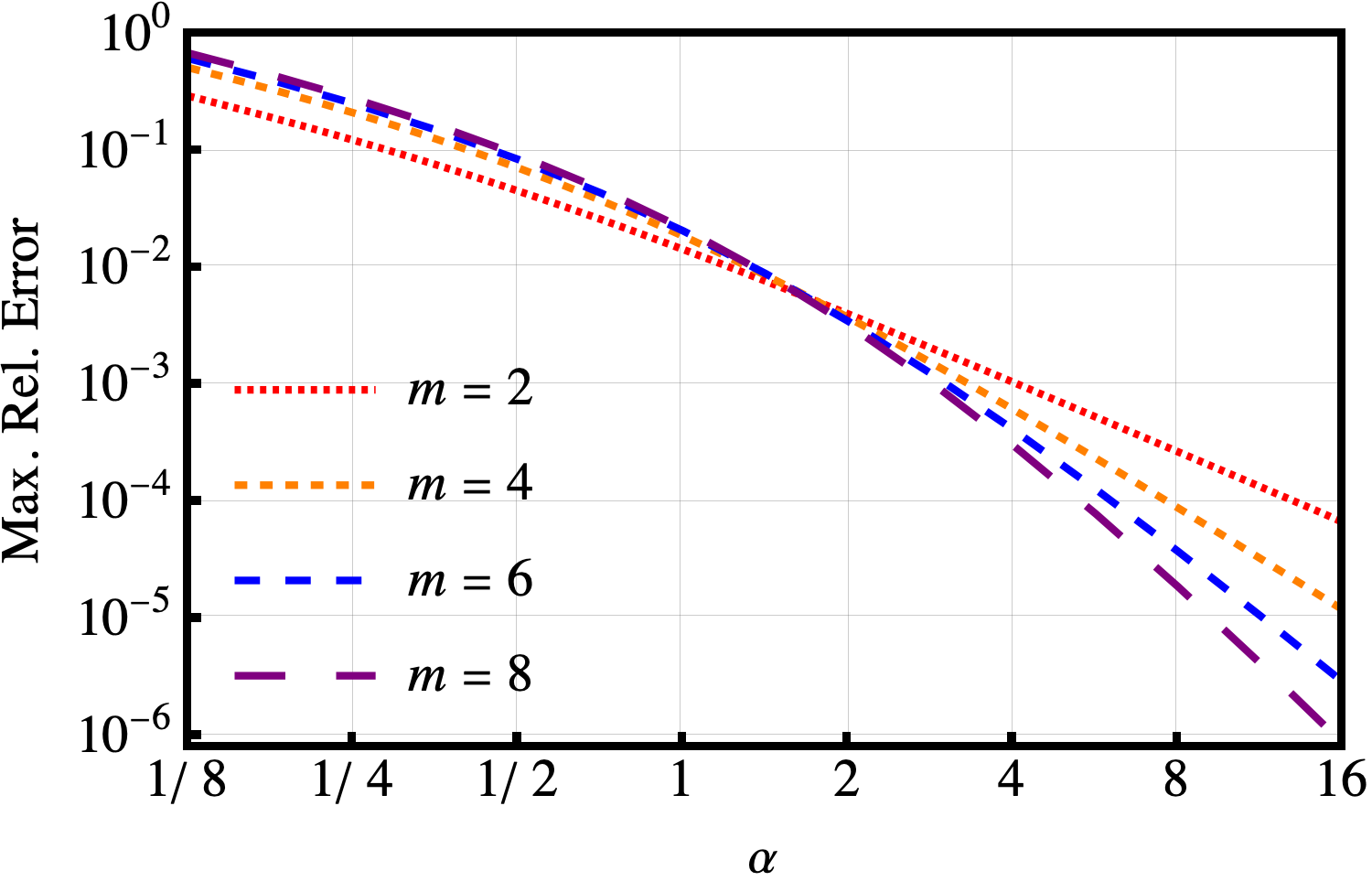}
\par\end{centering}
\caption{
The maximum relative error of $K_{\alpha}\left(x\right)_{\text{SCE}}^{\left(m\right)}$ as a function of $\alpha$ (note the x-axis is also logarithmic). For every $m$, the maximum error is seen to rapidly shrink with increasing $\alpha$ indicating that all the approximations are better for large $\alpha$. Additionally, one observes a cross-over around $\alpha\approx1.68$. For values of $\alpha$ larger than this, the maximum relative error decreases as the order $m$ is increased and thus increasing the order of the approximation uniformly improves the quality of the approximation. In contrast, for values of $\alpha$ smaller than this value, the maximum relative error increases and thus while higher orders may improve the accuracy of the large $x$ tail of $K_{\alpha}\left(x\right)_{\text{SCE}}^{\left(m\right)}$ (as seen in Fig.~\ref{fig:Bessel 1 SCE}), they cannot uniformly improve the approximation everywhere.
\label{fig:Max Rel Error}}
\end{figure}

Similar results are obtained for other values of $\alpha$ and, for the case of $m=2$, there is no noticeable difference between using Eq.~(\ref{eq:z1 Pade}) to approximate $z_1$ versus finding the root of Eq.~(\ref{eq:g(z1)}). Fig.~\ref{fig:Bessel alpha SCE}
shows the relative error, \mbox{$| (K_\alpha(x)^{(2)}_{\text{SCE}} - K_\alpha(x))/K_\alpha(x) |$}, of the second order SCE approximation $K_{\alpha}\left(x\right)_{\text{SCE}}^{\left(2\right)}$
for various $\alpha$. In both subfigures, the dotted red lines denote the relative error of
$K_{1}\left(x\right)_{\text{SCE}}^{\left(2\right)}$ and can be used
as a reference. Though the SCE is clearly seen to be able to approximate
$K_{\alpha}\left(x\right)$ for any $\alpha > 0$, it is also clear that
it performs far better for large $\alpha$ than for small $\alpha$.
Indeed, Fig.~\ref{fig:Max Rel Error} shows the maximum relative error of each approximation as a function of $\alpha$ and for every order $m$, we find that the relative error of $K_{\alpha}\left(x\right)_{\text{SCE}}^{\left(m\right)}$ is more tightly bounded as $\alpha$ is increased. This figure also makes it apparent that higher orders can in fact be used to uniformly improve the approximations but only if $\alpha$ is sufficiently large. For small $\alpha$, the lowest order approximation, $m=2$, is optimal for \textit{uniformly} bounding the relative error but this ceases to be the case once $\alpha$ exceeds roughly $\sim1.68$. Note however that if one is only interested in large values of $x$, then, as shown in Fig.~\ref{fig:Bessel 1 SCE}, higher orders can be used to obtain better approximations, even for small values of $\alpha$.

\begin{figure}
\begin{centering}
\includegraphics[width=0.49\textwidth]{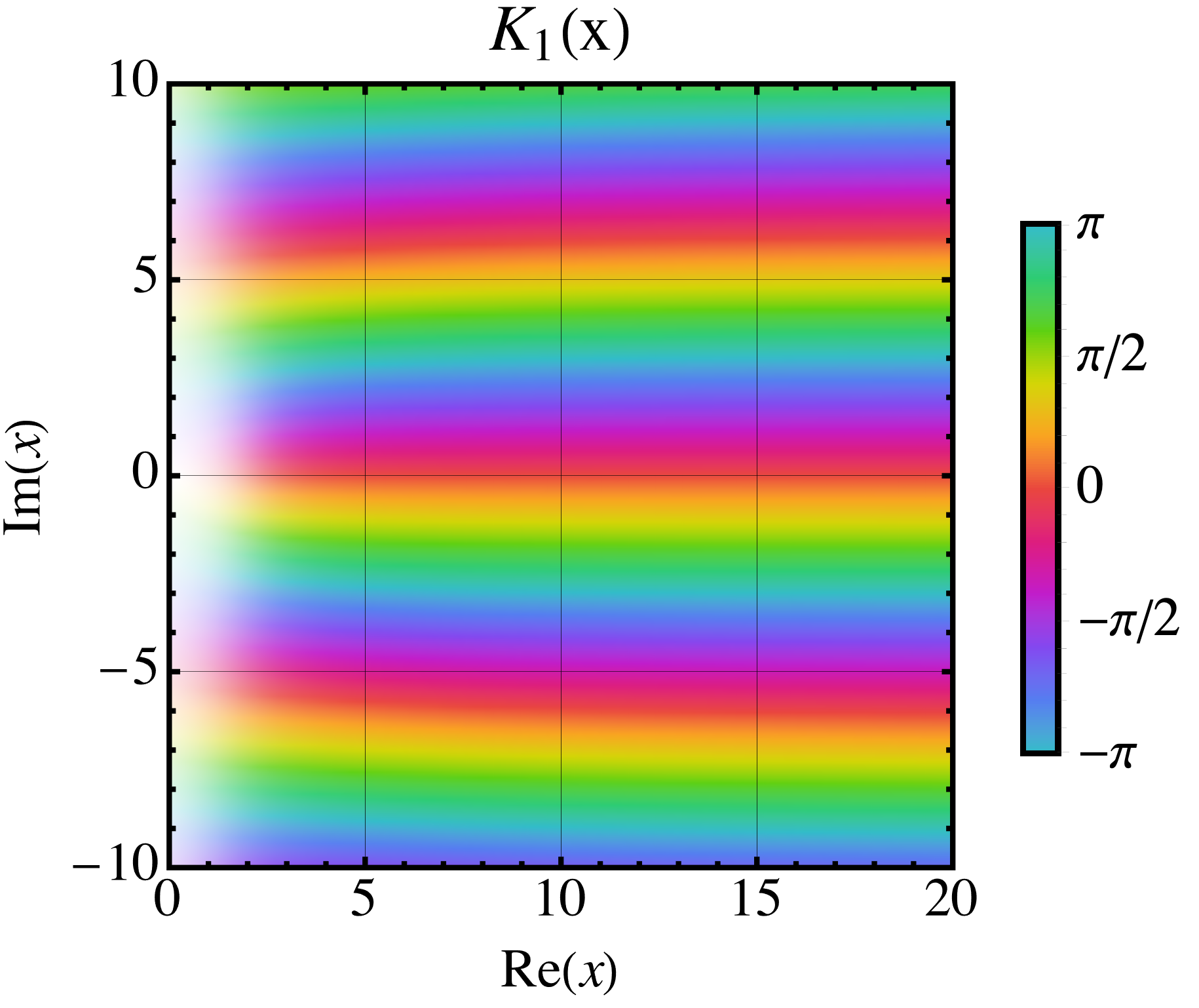}
\includegraphics[width=0.49\textwidth]{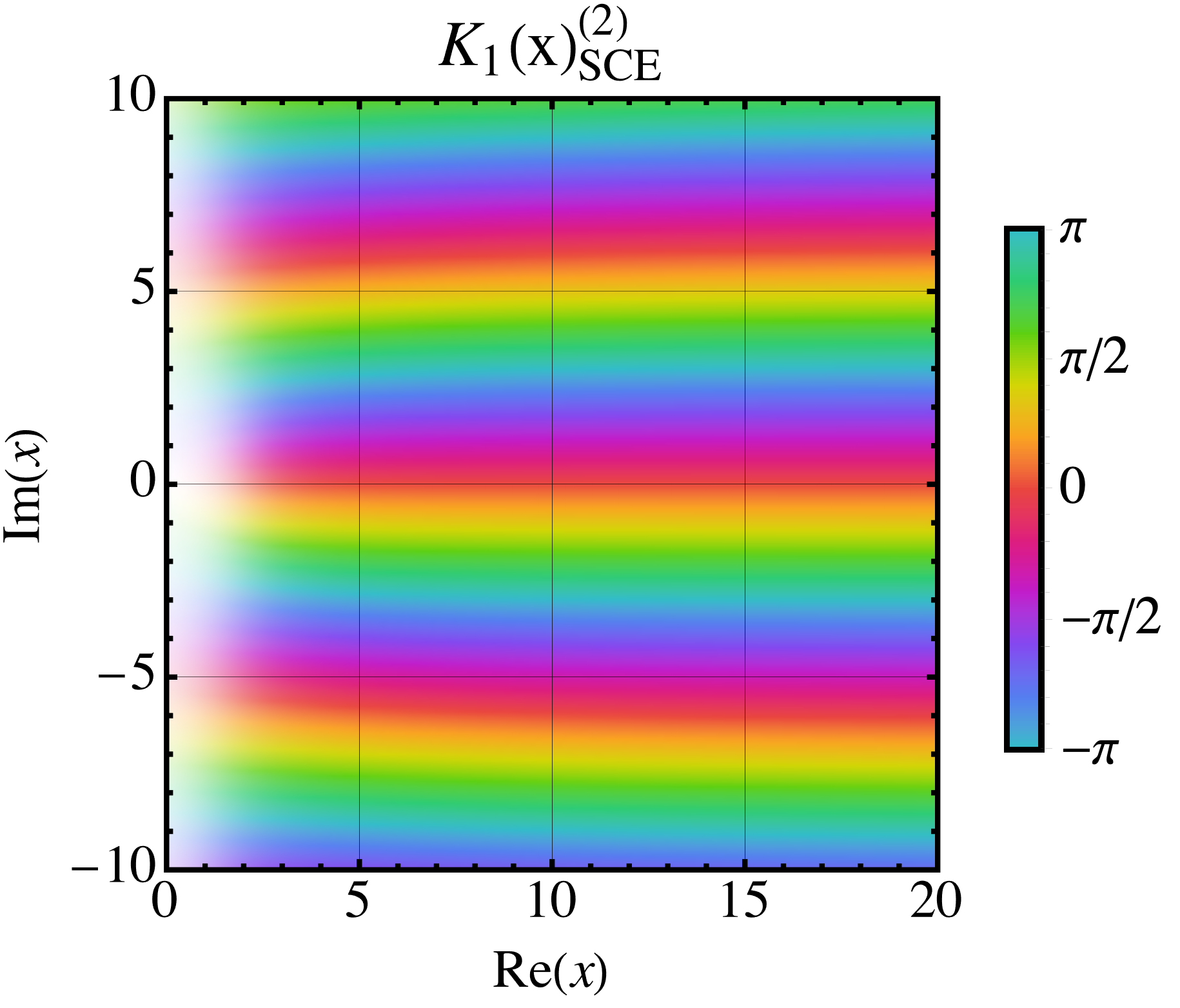}
\par\end{centering}
\vspace{0.5cm}

\begin{centering}
\includegraphics[width=0.5\textwidth]{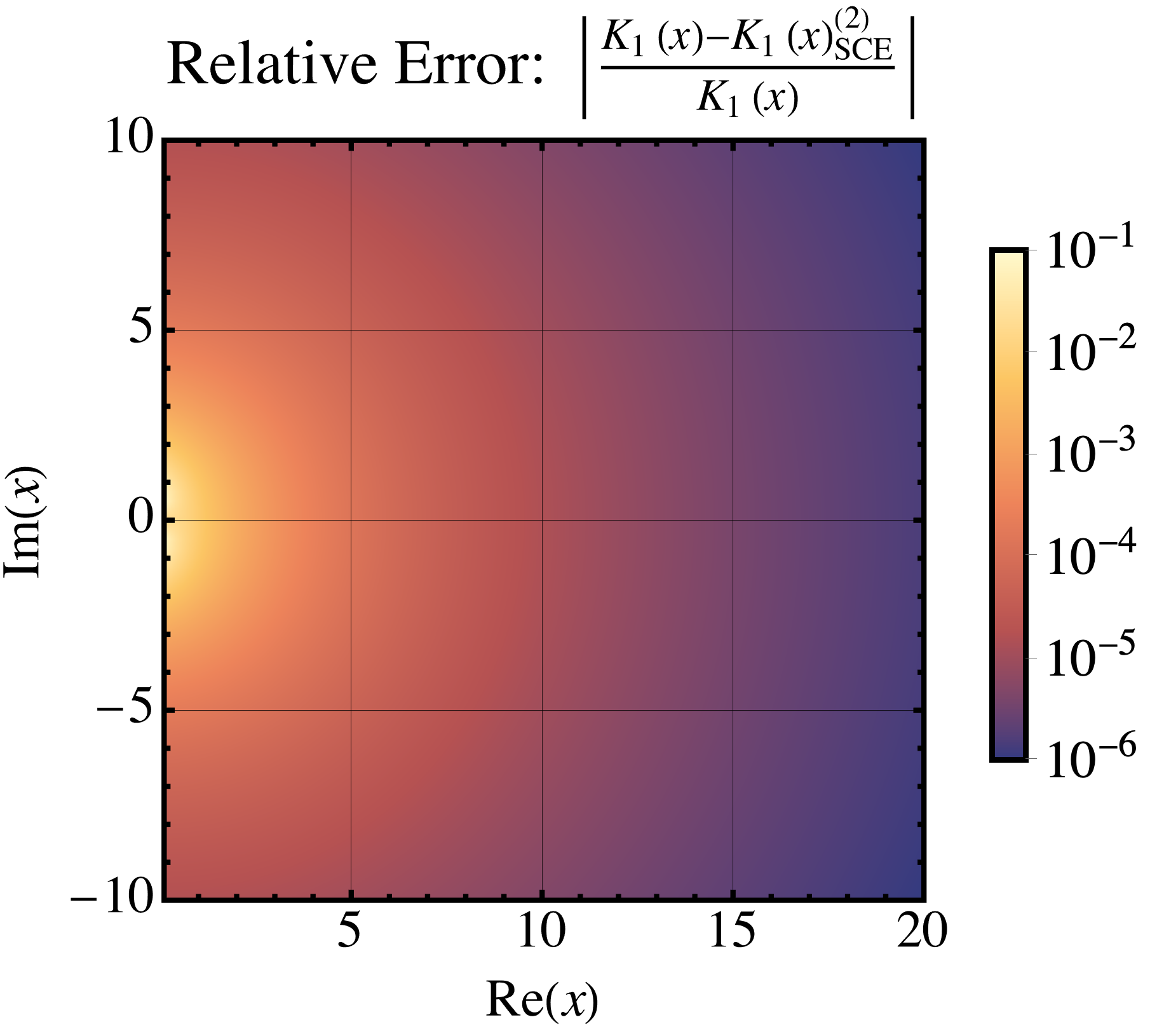}
\par\end{centering}
\caption{
$K_{1}\left(x\right)$ and $K_{1}\left(x\right)_{\mathrm{SCE}}^{\left(2\right)}$ as functions of $x\in\mathbb{C}$ (note that we only show the right half-plane $\mathrm{Re\left(x\right)>0}$). In the upper two figures, the hue corresponds to complex argument, $\arg K_{1}\left(x\right)$, while the brightness corresponds to its magnitude, 
$\left|K_{1}\left(x\right)\right|$. The low order approximation $K_{1}\left(x\right)_{\mathrm{SCE}}^{\left(2\right)}$ can be seen to qualitatively capture the behaviour of $K_{1}\left(x\right)$ over the entire right half-plane. The relative error between the approximation $K_{1}\left(x\right)_{\mathrm{SCE}}^{\left(2\right)}$ and $K_{1}\left(x\right)$ is shown in the lower figure and can be seen to rapidly decrease as one moves away from the origin in all directions. 
\label{fig:complex K}}
\end{figure}

Let us finish this section with a brief consideration of the convergence properties of our approximations on the complex plane. Though our primary goal has been to uniformly approximate $K_{\alpha}\left(x\right)$ for $x\in\mathbb{R}$, it is interesting and important to understand whether our approximation scheme generalises to complex values $x\in\mathbb{C}$. The power series and asymptotic expansions discussed in the introduction [Eqs.~(\ref{eq:I series}), (\ref{eq:K series 1}), (\ref{eq:K series 2}) and (\ref{eq:K large x series})] are in fact valid over the entire complex plane, with the exception of the negative real axis where a branch cut occurs, but the integral representation of $K_{\alpha}\left(x\right)$ given by Eq.~(\ref{eq:K integral}) only converges for $x\in\mathbb{C}$ when $\mathrm{Re}\left(x\right)>0$, ie. the right half-plane \cite{NIST:DLMF}. Since our entire approximation scheme is based on approximating this integral representation, it would be surprising for our approximation to be able to approximate $K_{\alpha}\left(x\right)$ outside of the right half-plane. Since the auxiliary functions $z_{1}\left(\alpha\right)$ and $\sigma_{0}^{2}\left(\alpha\right)$ only depend on $\alpha$ and the order of the approximation $m$, these quantities remain the same regardless of whether $x$ is real or complex. On the other hand, the function $\sigma^{2}\left(x,\alpha\right)$ given implicitly by Eq.~(\ref{eq:x(sigma)}) must necessarily become complex as $x$ acquires a non-vanishing imaginary part. Further, when we allow $x$ and $\sigma^{2}\left(x,\alpha\right)$ to take on complex values, Eq.~(\ref{eq:x(sigma)}) no longer becomes one-to-one and thus Eq.~(\ref{eq:x(sigma)}) can no longer be simply inverted to obtain $\sigma^{2}$ as a function of $x$. Despite having several solutions, Eq.~(\ref{eq:x(sigma)}) has only one solution that can be considered a continuous generalisation of the solution on the real axis and thus Eq.~(\ref{eq:x(sigma)}) can be solved numerically using initial values close to the solution for $\sigma^{2}\left(x,\alpha\right)$ on the real axis. While this works for all $x\in\mathbb{C}$ with $\mathrm{Re}\left(x\right)>0$, this solution ceases to be accessible once $\mathrm{Re}\left(x\right)<0$ and thus, as expected, our approximations are unable to describe $K_{\alpha}\left(x\right)$ in the left half-plane. 

Fig.~\ref{fig:complex K} shows a complex representation of $K_{1}\left(x\right)$ and the second order approximation $K_{1}\left(x\right)_{\mathrm{SCE}}^{\left(2\right)}$ over the right half-plane, as well as the relative error between them. In the top row, the hue corresponds to the complex arguments, $\arg K_{1}\left(x\right)$ and $\arg K_{1}\left(x\right)_{\mathrm{SCE}}^{\left(2\right)}$, while the brightness corresponds to the magnitudes, $\left|K_{1}\left(x\right)\right|$ and $|K_{1}\left(x\right)_{\mathrm{SCE}}^{\left(2\right)}|$. As can be seen, the approximation $K_{1}\left(x\right)_{\mathrm{SCE}}^{\left(2\right)}$ qualitatively captures the behaviour of $K_{1}\left(x\right)$ over the entire right half-plane. The plot in the second row shows the relative error between the exact function and the approximation and, like in the purely real case, we find that the relative error rapidly decreases as $\left|x\right|$ becomes large. Similar results are obtained for approximations $K_{\alpha}\left(x\right)_{\mathrm{SCE}}^{\left(m\right)}$ with higher orders $m$ and for other values $\alpha$. Accordingly, we find that even though our approximations were designed to approximate $K_{1}\left(x\right)$ for $x\in\mathbb{R}$, they are not particularly limited by such assumptions.

\FloatBarrier

\section{Discussion\label{sec:Discussion}}

We have shown that asymptotic matching of the SCE can produce superb
uniform approximations of the modified Bessel functions of the second
kind $K_{\alpha}\left(x\right)$. Indeed, it is reasonable to suspect
that this kind of asymptotic matching should be generically successful
and thus highly valuable whenever such asymptotic results are known.
The success of the application of a technique from statistical physics
to the task of approximating a ``special function'' also demonstrates
the substantial range of applicability of such methods.

It is worth drawing attention to the moment generating function formalism
used in this paper. Previous applications of the SCE have tended to
focus on imposing that the zeroth order moments $\left\langle u^{k}\right\rangle _{0}$
be exact up to some order \cite{Schwartz2008,Remez2018}. In principle,
we could have used the same approach and then attempted to modify
the moment order $k$ to obtain the desired asymptotic behaviour.
This however is challenging as $k$ is typically discrete and it is
unlikely that integer $k$ will manage to produce the desired asymptotic
behaviour. In \cite{Remez2018}, this limitation is circumvented
by first assuming that $k$ is an (even) integer and then simply generalising
the resultant expressions to any $k\in\mathbb{R}$. While it is unclear why
this should be legal, such an approach could certainly have been used
here though the additional complication introduced by requiring two
such constraints for the two free parameters $\xi$ and $\sigma$
would make the equations substantially more involved than in that
case. Instead, we have chosen to introduce a new formalism of focusing
on the moment generating function $\left\langle e^{uz}\right\rangle $.
In particular, we have attempted to impose that the zeroth order approximation
of this function of $z$ be exact up to first order. Since
\begin{equation}
\left\langle e^{uz}\right\rangle _{0}=\sum_{k=0}^{\infty}\frac{1}{k!}\left\langle u^{k}\right\rangle _{0}z^{k}\,,
\end{equation}
imposing that this expression be exact up to first order at some point
$z_{1}$ is equivalent to imposing that a particular linear combination
of all the moments $\left\langle u^{k}\right\rangle _{0}$ be exact
up to first order. The task of deciding which linear combination to
select can then be performed by demanding the desired asymptotic matching.
It is important to appreciate how this approach is simultaneously
simpler and less arbitrary than the previously used approaches of
directly targeting the moments $\left\langle u^{k}\right\rangle $.
Additionally, as pointed out just before Eq.~(\ref{eq:moment constraints}),
the resultant equations for the moment generating function $\left\langle e^{uz}\right\rangle $
can easily be converted into equations for the moments $\left\langle u^{k}\right\rangle $
by mere repeated differentiation. Accordingly, nothing is lost by
first considering the expansion of the moment generating function,
even if one ultimately settles on using the moments themselves.

Despite its great success, there is also much to learn from the limitations
of the method. The most glaring limitation of our results is that
for small $x$, taking higher orders provides no additional benefit
over the lowest order approximation when $\alpha\lesssim1.68$. 
That is, the asymptotic matching
provides superb approximations but, at least under these circumstances, the quality
of these approximations is ``frozen'' for small $x$. 
The origin of this behaviour is undoubtedly the fact that in the limit of $x \rightarrow 0$,
the Hamiltonian given by Eq.~(\ref{eq:Bessel Hamiltonian}) ceases to be bounded and 
thus our ``partition function'', given by Eq.~(\ref{eq:K integral}), becomes singular.
As such, while we suspect that this is not a generic feature of the method, it is also likely
not unique to this problem. Accordingly, for problems which have this
frozen characteristic, additional techniques need to be introduced
to ``unfreeze'' the quality of the approximations. One very natural
suggestion is to use Eq.~(\ref{eq:sum of I_n}) to ensure that the
zeroth order approximation for $K_{\alpha}\left(x\right)$ is exact
up to say $m^{th}$ order instead of just first order. Alternatively,
Eq.~(\ref{eq:mean expansion}) could be used to ensure that the zeroth
order moment generating function $\left\langle e^{uz}\right\rangle _{0}$
be exact up to higher orders. While both of these suggestions are
likely to succeed at unfreezing the quality of the approximations,
they suffer from the original problem that asymptotic matching was
supposed to solve, namely too much freedom! As before, it is not a
priori obvious which of these new approximations will succeed and
thus one must again resort to comparison with empirical results. We
thus consider such a resolution to be unsatisfactory. Developing an
approach capable of unfreezing such approximations without resorting
to comparisons with empirical results is thus left as an open question
for study.

Finally, in this paper, we have made use of an interesting comparison
between an integral representation of the mathematical function $K_{\alpha}\left(x\right)$
and the partition function $Z$ describing the equilibrium properties
of a single particle in a certain hyperbolic potential-well to develop
fruitful insights into $K_{\alpha}\left(x\right)$. As a purely mathematical
object, it is interesting to ask whether other insights from statistical
physics could shed light on the properties of $K_{\alpha}\left(x\right)$.
For instance, can the dynamical properties of a system governed by
the Hamiltonian defined in Eq.~(\ref{eq:Bessel Hamiltonian}) tell
us anything interesting about the function $K_{\alpha}\left(x\right)$?
This too, we leave as an open question for contemplation.

\printbibliography

\end{document}